\documentclass[aip,pop,preprint]{revtex4-1}
\usepackage{graphicx}
\usepackage{amsmath}
\usepackage{amssymb}
\usepackage{bm}
\usepackage{natbib}
\usepackage{color}
\usepackage[normalem]{ulem}

\newcommand{\ue}{\ensuremath{\mathbf{u}_{_E}}}

\newcommand{\myemail}{jdahlin@umd.edu}

\begin{document}

\preprint{AIP/123-QED}


\title{Electron acceleration in three-dimensional magnetic reconnection with a guide field}

\author{J. T. Dahlin}
\email[]{\myemail}
\author{J. F. Drake}
\altaffiliation{Department of Physics, University of Maryland, College Park, Maryland 20742, USA}
\altaffiliation{Institute for Physical Science and Technology, University of Maryland, College Park, Maryland 20742, USA}
\altaffiliation{Space Science Laboratory, University of California, Berkeley, California 94720, USA}

\author{M. Swisdak}

\affiliation{Institute for Research in Electronics and Applied Physics, 
University of Maryland, College Park, Maryland 20742, USA}

\begin{abstract}

Kinetic simulations of 3D collisionless magnetic reconnection with a 
guide field show a dramatic enhancement of 
energetic electron production
when compared with 2D systems. In the 2D systems, electrons are trapped 
in magnetic islands that limit their energy gain, whereas in the 3D 
systems the filamentation of the current layer leads to a stochastic 
magnetic field that enables the electrons to access volume-filling 
acceleration regions. The dominant accelerator of the most energetic 
electrons is a Fermi-like mechanism associated with reflection of 
charged particles from contracting field lines.

\end{abstract}



\maketitle


Magnetic reconnection is a ubiquitous plasma process that converts
magnetic energy into thermal and kinetic energy. Of particular
interest is the production of non-thermal particles, in which a
fraction of the plasma population is driven to energies much larger
than that found in the ambient medium. Reconnection is thought to be an
important driver of such particles in phenomena such as gamma ray bursts
\citep{drenkhahn02a,michel94a}, stellar and solar flares
\citep{lin03a}, and magnetospheric storms \citep{oieroset02a}.  Recent
observations of solar flares reveal the remarkable efficiency of
electron acceleration: a large fraction of the electrons in the flaring
region become a part of the nonthermal spectrum, with a resulting
energy content comparable to that of the magnetic field
\citep{krucker10a,oka13a}.

Mechanisms for particle acceleration have been explored and compared
in a variety of papers,
e.g. \cite{hoshino01a,zenitani01a,drake05a,pritchett06a,egedal09a,oka10a,hoshino12a}.
Several authors
\citep{litvinenko96a,drake05a,egedal12a} have examined acceleration by
electric fields parallel to the local magnetic field ($E_\parallel$). However, 
parallel electric fields are typically localized near reconnection
X-lines and separatrices, which limits the number of electrons that can be accelerated through this mechanism.

Drake et al. \cite{drake06a} proposed a mechanism whereby
charged particles gain energy as they reflect from the ends of 
contracting magnetic islands, a process analogous to the first-order 
Fermi acceleration of cosmic rays. This mechanism operates wherever
there are contracting field lines in a reconnection region, and therefore
develops during single X-line reconnection or as magnetic islands merge 
\citep{drake06a,oka10a,drake10a,drake13a,hoshino12a}. 
This mechanism is therefore volume filling and can accelerate a large number 
of particles. 

In a recent article \cite{dahlin14a}, we developed a method for 
calculating electron acceleration due to three fundamental mechanisms:
parallel electric fields $E_\parallel$, betatron acceleration associated
with conservation of the magnetic moment, and Fermi reflection due to the
relaxation of curved magnetic field lines.
We found that Fermi reflection dominated in reconnection where the magnetic 
fields are roughly antiparallel (see also Guo et al. \citep{guo14a}),
whereas in guide field reconnection 
Fermi reflection
and $E_\parallel$ were both important drivers of particle acceleration.
However, we did not address the scaling of each mechanism with particle energy. 
This is important for determining the mechanism responsible for producing the 
most energetic particles.

Studies of particle acceleration in reconnection have primarily been based 
on 2D simulations, 
in which accelerated particles are typically localized near the X-line, along 
magnetic separatrices and within magnetic islands \citep{drake05a,guo14a}. There 
are some observations with small ambient guide fields \citep{chen07a,retino08a,huang12a} 
that support such a picture. A notable exception are Wind observations 
in which energetic electrons up to 300 keV are seen for more than an hour in an 
extended region around the reconnection region \citep{oieroset02a}. These observations 
correspond to reconnection with a strong guide field. 

Two-dimensional simulations impose limitations on the magnetic topology as well as 
the available spectrum of instabilities. 
In the presence of an ambient guide 
field, 
reconnection in 3D can become 
turbulent as a result of the generation of magnetic islands along 
separatrices and adjacent surfaces \cite{schreier10a,daughton11a}. While 
test particle trajectories in MHD fields have been used to explore acceleration 
in such systems, \citep{onofri06a,kowal11a} the absence of feedback of energetic 
particles on the reconnection process in such models limits their applicability 
to real systems. Recent 3D studies of kinetic reconnection examined particle 
acceleration in electron-positron plasmas \citep{sironi14a,guo14a}. 
However, these studies focused on relativistic regimes where the magnetic energy 
per particle exceeds the rest mass energy and included no ambient guide field.
Hence, the impact of complex 3D magnetic fields on particle acceleration 
remains an open topic.

Here, we explore magnetic reconnection in 3D systems with a strong guide field
(i.e. one sufficiently strong to magnetize the electrons and to play
an important role in pressure balance),
which is the most generic form of reconnection in space and astrophysical plasmas.
We find that
the efficiency of particle acceleration is greatly increased compared to that in
2D systems (where $d/dz = 0$, as in Dahlin et al.\cite{dahlin14a}). We show that this occurs because 
the complex 3D magnetic fields enable 
the most energetic particles to continually access volume-filling acceleration sites rather than 
being confined to a single magnetic island that no longer accelerates particles once it 
has fully contracted. We also examine the energy dependence of the dominant 
$E_\parallel$ and Fermi acceleration mechanisms, and find that Fermi reflection is 
the primary accelerator of the energetic electrons.

We explore particle acceleration via simulations using the massively parallel 3D
particle-in-cell (PIC) code {\tt p3d} \citep{zeiler02a}. Particle trajectories
are calculated using the relativistic Newton-Lorentz equation, and the electromagnetic
fields are advanced using Maxwell's equations. The time and space coordinates are 
normalized, respectively, to the proton cyclotron time $\Omega_{ci}^{-1} = m_i c/eB$ 
and inertial length $d_i = c/\omega_{pi}$. 
The grid cell width is $d_e/4$, where $d_e = d_i \sqrt{m_e/m_i}$ is the electron
inertial length. 
The time step is $dt = 0.01 \Omega_{ci}^{-1} = 0.25 \Omega_{ce}^{-1}$, where
$\Omega_{ce} = (m_i/m_e)\Omega_{ci}$ is the electron cyclotron frequency.

We focus on a 3D simulation with dimensions 
$L_x \times L_y\times L_z = 51.2 d_i\times 25.6d_i \times 25.6d_i$ and 
an analogous 2D simulation with $L_x \times L_y = 51.2 d_i \times 25.6 d_i$. 
These simulations use an artificial proton-to-electron mass 
ratio $m_i/m_e = 25$ in order to reduce the computational expense.
Simulations with differing mass-ratios and domains are presented 
to demonstrate the generality of the results.

All simulations are initialized with a force-free configuration and use
periodic boundary conditions. This is chosen as the most generic model
for large-scale systems such as the solar corona where the density jump
between the current layer and upstream plasma is not expected to be important.
The magnetic field is given by:
$B_x = B_0 \tanh(y/w_0)$ and $B_z = \sqrt{2B_0^2-B_x^2}$, corresponding to
an asymptotic guide field $B_z\infty = B_x\infty = B_0$.
We include two current sheets at $y=L_y/4$ and $3L_y/4$ to produce a periodic
system, and $w_0 = 1.25d_e$.
This initial configuration is not a kinetic equilibrium, which
would require a temperature anisotropy \cite{bobrova01a}, but is in pressure
balance.

The 3D simulations use at least 50 particles per cell for each species, and the 2D
simulations use $1600$ particles per cell. 
The initial electron and proton temperatures are isotropic, with $T_e = T_i = 0.25m_i c_A^2$, 
and the initial density $n_0$ and pressure $p$ are constant so that $\beta = 8\pi p/B^2 = 0.5$. 
The speed of light is $c = 3 c_A \sqrt{m_i/m_e}$, where $c_{A}=B_0/\sqrt{4\pi m_i n_0}$.

\begin{figure}
 \includegraphics{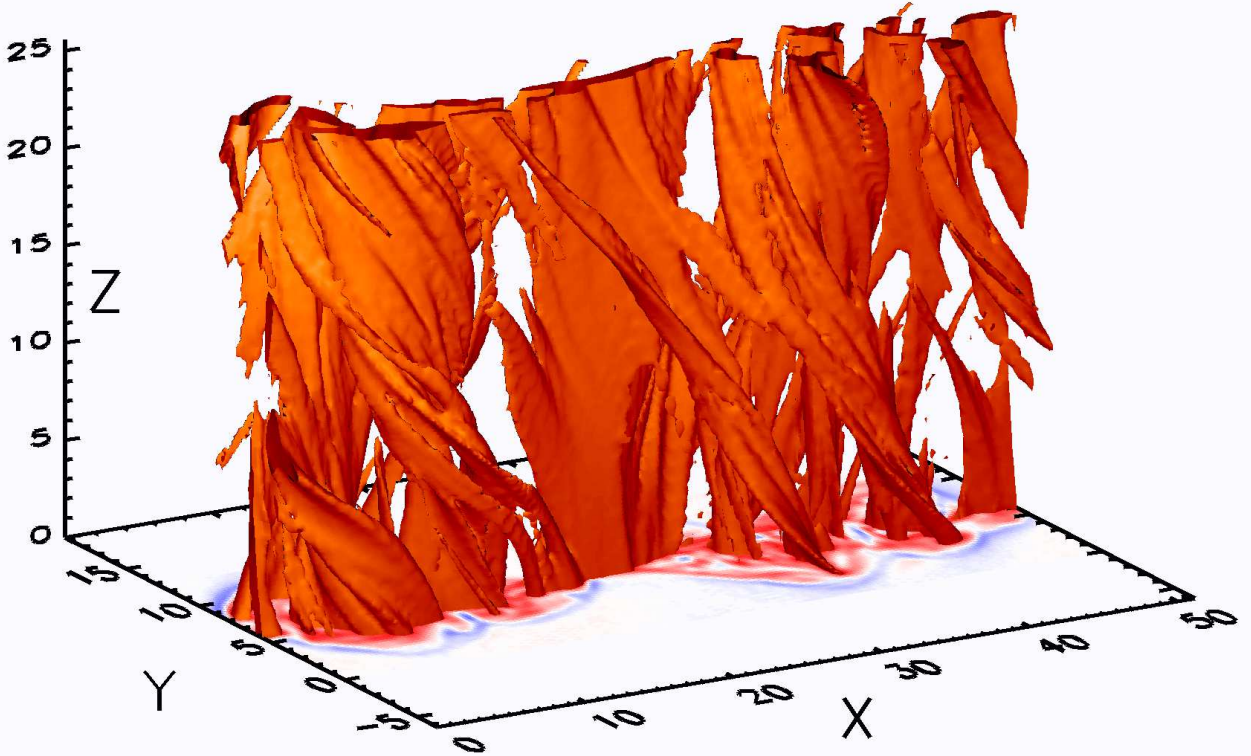}
 \caption{Isosurface of $J_{ez}$ at $t \Omega_{ci} = 50$. The
isosurface level is 60\% of the maximum current density (a 2D slice
of the same quantity is shown on the bottom). 
\label{fig:p15isojez}}
 \end{figure}

Reconnection develops from particle noise via the tearing instability, generating
interacting flux ropes that grow and merge until they reach the system size at
$t\Omega_{ci} \sim 50$. The macroscopic evolution of the 2D and 3D
systems is similar at this point, though the 2D simulation has released roughly 15\% more magnetic
energy.  Fig. \ref{fig:p15isojez} shows an isosurface of one component
of the electron current density $J_{ez}$ 
at $t\Omega_{ci} = 50$ in the 3D simulation. The current exhibits filamentary 
structure that develops from instabilities with $k_z \neq 0$ 
that are prohibited in 2D reconnection simulations \cite{daughton11a}.

\begin{figure}
 \includegraphics{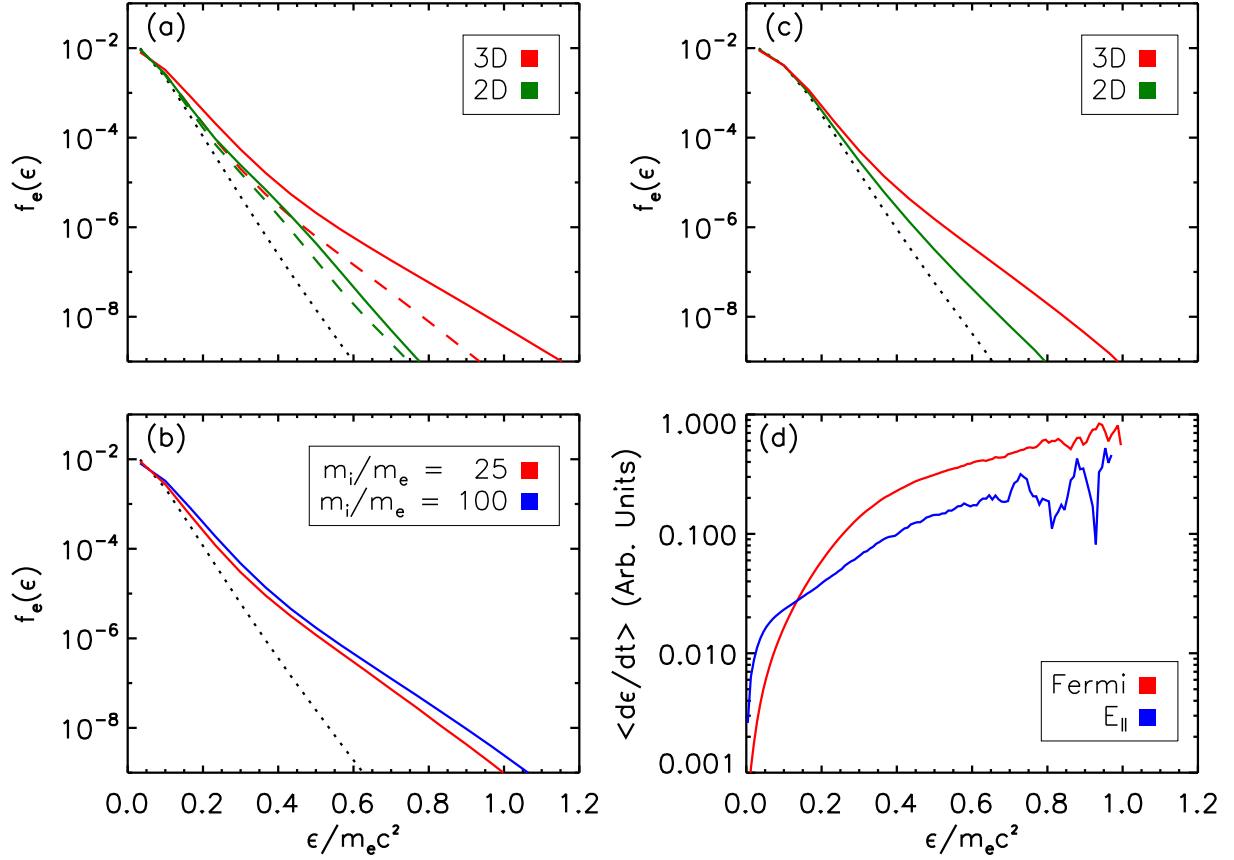}
 \caption{
(a-c) Global electron energy spectra. 
The 3D simulation dimensions $L_x \times L_y \times L_z$ are: 
(a) $102.4 \times 51.2 \times 25.6$,
(b) $51.2 \times 25.6 \times 12.8$,
(c) $51.2 \times 25.6 \times 25.6$.
Dotted lines indicate initial spectra, 
solid lines in (b-d)
correspond to $t=50$.
Dashed and solid lines in (a) correspond to 
$t=50$ and $t=125$, respectively. 
(d) Average electron energization rate
vs. energy for the 3D simulation shown in (c).
\label{fig:p15allspectra}}
 \end{figure}

In Fig. \ref{fig:p15allspectra}, energy spectra are shown for a variety of simulations 
in 2D and 3D with differing domain sizes and mass ratios. The spectra reveal 
significant electron acceleration in both 2D and 3D simulations. However, the 3D 
simulations produce a much greater number of energetic particles: the fraction
of electrons with energy exceeding $0.5 m_e c^2$ is roughly an order of magnitude larger
than in the 2D simulations ($\sim 2 \times 10^{-4}$ vs. $\sim 2 \times 10^{-5}$).
The separation between the spectra in the 2D and 3D systems
is greatest in the largest simulation domain (compare panels (a) and (c)) suggesting that 
this difference would be even greater for larger systems.
Since the magnetic energy dissipation is typically slightly larger in the 2D systems, 
the increased energetic electron production in the 3D systems is due to enhanced
acceleration efficiency rather than an increase in the total energy imparted to
the plasma. These results are insensitive to the mass ratio (panel (b)).
The discussion that follows will focus on the results of the simulations shown in
panel (c).

\begin{figure}
 \includegraphics{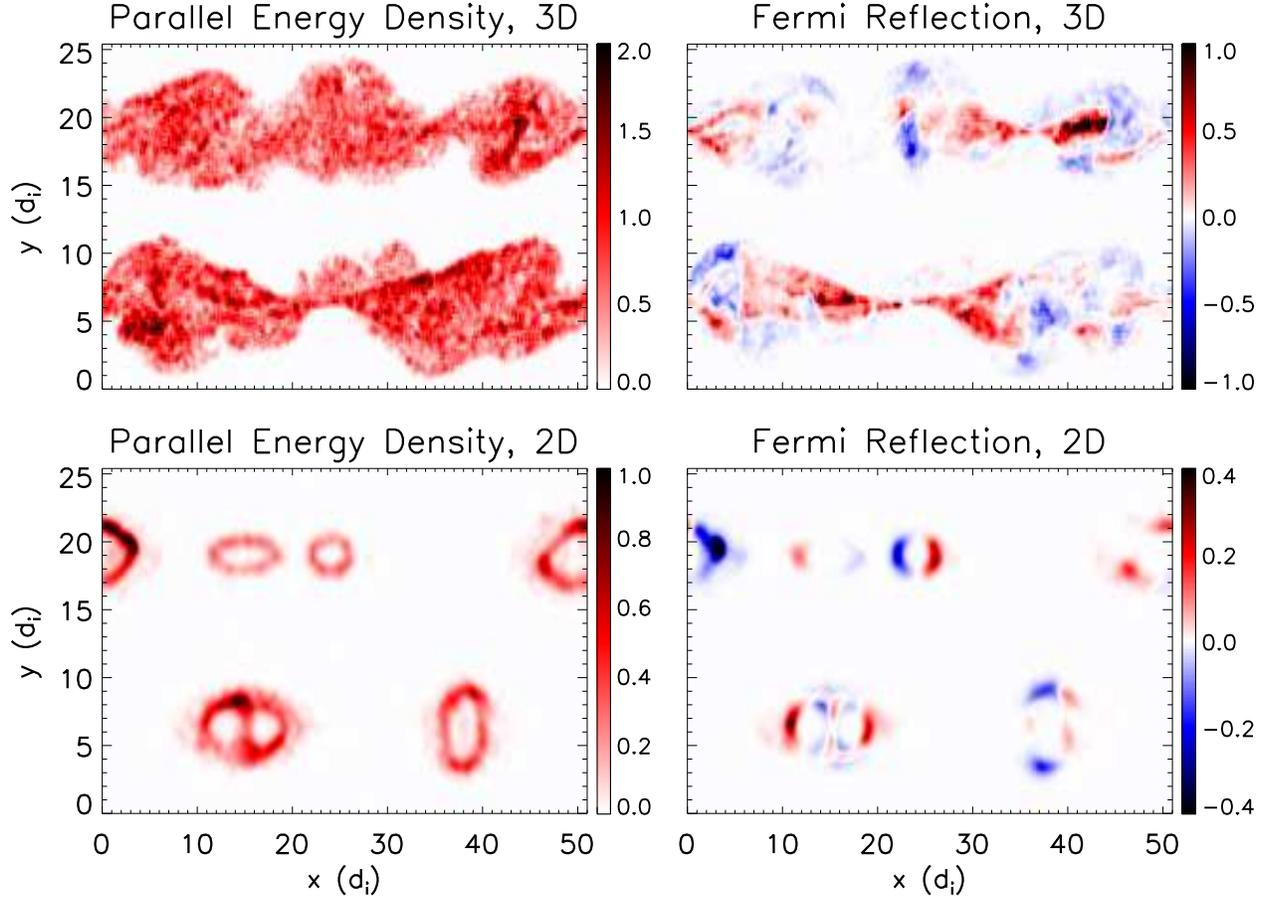}
 \caption{
Parallel energy density and Fermi reflection heating rate in 2D and 3D
for electrons with $\epsilon > 0.5 m_ec^2$ in the plane $z=0$ at $\Omega_{ci}t = 50$.
\label{fig:p15heat3d_50}}
 \end{figure}

The spatial distribution of the most energetic particles (shown in the left-hand panels of Fig. \ref{fig:p15heat3d_50})
also differs between the 2D and 3D simulations: these particles occupy narrow
bands well inside the islands
in the 2D simulation, but are distributed throughout the reconnecting region in the
3D simulation. 
In the 2D system, the reconnected field lines form closed loops (islands) that trap 
particles. 
The stochastic structure of the magnetic field in the 3D system, however, allows 
field-line-following particles to wander throughout the chaotic reconnecting region 
\cite{jokipii69a}. Surfaces of section from field line tracing reveal that the 
region of stochastic magnetic field in the 3D simulation roughly matches the
distribution of energetic electrons.
The distribution of the energetic particles in the 3D simulation 
is broadly consistent with Wind magnetotail observations where energetic 
electrons are seen for more than an hour and therefore must occupy 
a large region \cite{oieroset02a}.

In order to examine the mechanisms responsible for accelerating these particles, 
we assume a guiding-center approximation relevant for a 
strong guide field \citep{northrop63a}. In this limit, the evolution of the kinetic energy $\epsilon$ 
of a single electron can be written as:
\begin{equation}
\label{eqn:particle}
\frac{d \epsilon}{d t} = q E_\parallel v_\parallel
  + \frac{\mu}{\gamma}\left( \frac{\partial B}{\partial t} + \ue \boldsymbol{\cdot} \boldsymbol{\nabla} B \right)
  + \gamma m_e v_\parallel^2 (\ue \boldsymbol{\cdot} \boldsymbol{\kappa})
\end{equation}
where $E_\parallel = \mathbf{E} \boldsymbol{\cdot} \mathbf{b}$ is the parallel electric field,
$\mu = m_e \gamma^2 v_\perp^2/2B$ is the magnetic moment, $\ue$ is the $\mathbf{E} \times \mathbf{B}$ velocity
corresponding to the advection of the magnetic field, 
and $\boldsymbol{\kappa} = \mathbf{b} \boldsymbol{\cdot} \nabla \mathbf{b}$ is the magnetic curvature.
The velocity components parallel and perpendicular to the magnetic field are
$v_\parallel$ and $v_\perp$, respectively; $\gamma$ is the relativistic Lorentz factor, and
$\mathbf{b}$ is the unit vector in the direction of the local magnetic field.

The first term on the right-hand-side of the equation corresponds to acceleration
by parallel electric fields, which are typically localized near the reconnection 
X-line and along separatrices. The second term corresponds to betatron acceleration,
which is typically negligible in reconnection \cite{dahlin14a}.
The last term corresponds to reflection of particles from contracting magnetic field lines,
a type of first-order Fermi acceleration \citep{drake06a,drake10a,hoshino12a}. This occurs where
tension is released as magnetic fields advect in the direction of magnetic curvature 
($\ue \cdot \boldsymbol \kappa > 0$).
Both $E_\parallel$ and Fermi reflection change the parallel energy 
of the particles, while betatron acceleration changes the perpendicular energy.

Equation (\ref{eqn:particle}) reveals that the acceleration mechanisms have 
different scalings with the particle energy:
the Fermi reflection term is second-order in the parallel velocity, 
whereas the parallel electric field term is only first-order. 
Panel (d) of Fig. \ref{fig:p15allspectra} shows the average acceleration 
per particle for both $E_\parallel$ and Fermi reflection in the
3D simulation at $t\Omega_{ci} = 50$. The bulk thermal electrons (low energies) 
are primarily accelerated by $E_\parallel$, whereas Fermi reflection
is more important at high energies,
consistent with the energy scaling of Eq. (\ref{eqn:particle}).

The spatial distribution of the Fermi reflection term for the most energetic electrons ($> 0.5 m_ec^2$)
is shown on the right-hand side of Fig. \ref{fig:p15heat3d_50}. While
acceleration occurs throughout the reconnection exhaust in the 3D simulation, in 2D
the acceleration is limited to narrow bands near the cores of magnetic islands. 
This contrast suggests that the stochastic 3D field structure allows the electrons 
to have greater access to the acceleration regions where magnetic energy is being released.

\begin{figure}
 \includegraphics{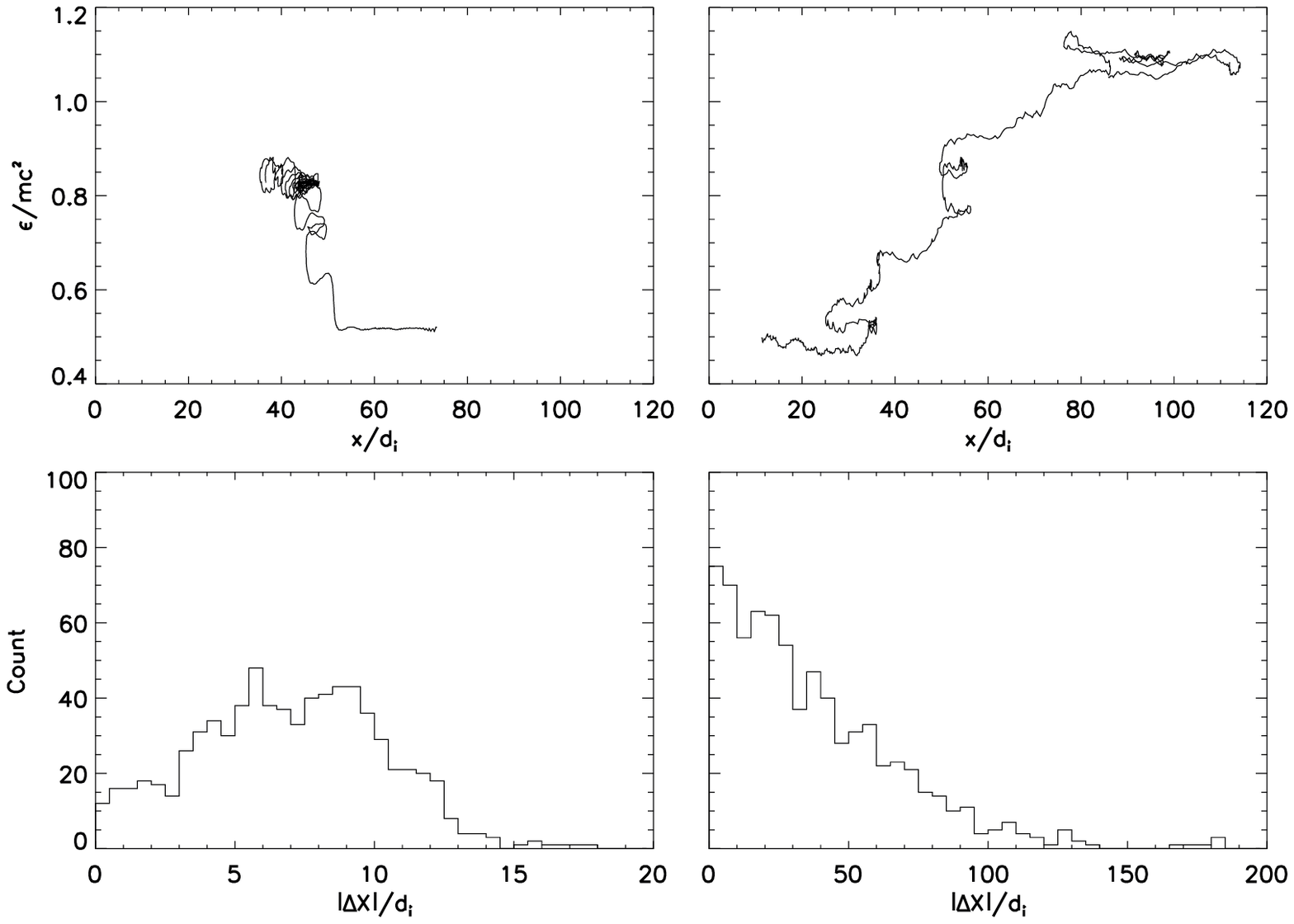}
 \caption{[Top] Typical energy vs. position plots for an energetic particle 
in the 2D system (left) and the 3D system (right) 
over the period
$\Omega_{ci}t = 0$ to $50$.
[Bottom] Distribution of $\Delta X = |x(t=50)-x(t=25)|$ for the 750 most
energetic particles in the 2D simulation (left) and 3D simulation (right). 
\label{fig:p15trajs}}
 \end{figure}

To explore the reason for enhanced acceleration in the 3D system, we examine the 
trajectories of the 750 most energetic electrons in each simulation. A typical
trajectory from the 2D simulation is 
shown in the top left panel in Fig. \ref{fig:p15trajs}. The electron begins in the
tail of the electron distribution with kinetic energy $\epsilon \approx 0.4 m_ec^2$. The
electron streams along a field line outside the reconnection region before accelerating 
at an X-line near $x \sim 50$ and becoming trapped in an island. The electron bounces 
several times inside this island, accelerating up to $\epsilon \approx 0.8 m_ec^2$. By this point,
the field line the electron is following has released its tension, so acceleration ceases
even as the electron continues to bounce.

The top right panel of Fig. \ref{fig:p15trajs} shows a typical electron trajectory from
the 3D simulation.
The electron moves throughout the reconnecting domain, and does not become trapped
 on stagnant field lines in island cores as in the 2D simulation. It instead gains energy at many different 
 acceleration sites, eventually reaching a maximum energy of $\epsilon \approx 1.15 m_ec^2$.
The Supplemental Material \citep{p15supplemental} includes videos of both particle trajectories
in the time-evolving field of the simulation.

The electron trajectories shown here are generic for their respective simulations.
Though the acceleration details differ, all of the electrons in the 2D simulation are
confined to single islands, whereas no electrons in the 3D simulation show significant trapping.
The bottom panels of Fig. \ref{fig:p15trajs} show the distribution of the particle displacement
$|\Delta x| = |x(\Omega_{ci}t=50)-x(\Omega_{ci}t=25)|$ for the 750 most energetic
particles in each simulation (the choice of $\Omega_{ci}t = 25$ as the earliest time 
eliminates free streaming along unreconnected field lines before islands develop). 
The average displacement of the energetic electrons in the 3D system is nearly an order
of magnitude greater than that in the 2D simulation, underscoring a fundamental 
difference in the particle trajectories of the two systems.

It has been shown previously that the development of pressure anisotropy with 
$P_\parallel \gg P_\perp$ causes the cores of magnetic islands to approach firehose 
marginal stability, where the tension driving magnetic reconnection ceases, thereby 
throttling reconnection. A full treatment of the feedback from particle acceleration 
(e.g. Drake et al.\cite{drake13a}) is outside the scope of this paper. However, we do find that 
large anisotropies persist in the 3D simulations 
(Supplemental Material \cite{p15supplemental}), 
so the turbulent dynamics do not appear to significantly isotropize the pressure.
The firehose instability is not triggered in these simulations, likely due to the 
strong ambient guide field - there is insufficient magnetic free energy to
reach $\beta \sim 1$. In cases with weaker guide fields the firehose feedback is
likely to be more important \cite{drake12a}.


The electron spectra in both simulations do not assume a power law
as is frequently observed in nature. This is due in part to the limited energy
gain possible in the modest-sized 3D simulation presented here. Previous 2D 
simulations have shown the total energy gain is greater in larger systems \cite{dahlin14a}. 
An additional issue is that these
simulations have periodic boundary conditions so no particles are lost from the
system. 
Solar observations suggest that electrons are confined in regions of energy release 
in the corona \citep{krucker10a}. The mechanism for confinement remains an open issue,
although both magnetic 
mirroring and double layers are possible mechanisms \citep{li12a}. On the other hand, 
it has been suggested that the development of a power law requires
a loss mechanism in addition to an energy drive \cite{drake13a}.
However, recent electron-positron simulations  \cite{guo14a,sironi14a}
suggest that power-law spectra may still develop in the absence of a loss mechanism. 
The set of conditions under which power law spectra form in kinetic reconnection
simulations remains an open issue.

A limitation of the present simulations is the use of an artificial mass ratio, which
reduces the separation between proton and electron scales. In order for an electron 
to access multiple acceleration sites as we observe in our simulations, its characteristic 
velocity must exceed that of the macroscopic flows associated with reconnection, which is
controlled by the protons. The results
of a simulation performed with $m_i/m_e = 100$ were not qualitatively different from
the $m_i/m_e = 25$ simulation, suggesting we have achieved a sufficient separation of
scales (see Fig. 2b).
The absence of such a separation of scales may explain why this behavior is not 
observed in electron-positron simulations \citep{guo14a}.

The plasma parameters used in these simulations are not tuned to a specific astrophysical
system. 
For example, the ratio of the speed of light to the electron Alfv\'en velocity
($c/c_{Ae} = 3$) 
is smaller than is typical for either the magnetotail \cite{oieroset02a} or the solar wind \cite{phan06a}. However,
so long as $c/c_{Ae} > 1$, the important reconnection dynamics are not relativistic.
The initial
temperature ($T_{e0} = m_i c_A^2/8$) is much larger than in the either the corona \citep{krucker10a} 
or the magnetotail \citep{oieroset02a}. As long as the electron thermal velocity $v_{te0} \gg c_A$, 
electrons circulate in islands faster than the islands contract and a preheating mechanism is not 
needed for Fermi reflection to function.

%



\begin{acknowledgments}
This work has been supported by NSF Grants
AGS1202330 and PHY1102479, and NASA grants
NNX11AQ93H, APL-975268, NNX08AV87G, NAS 5-
98033, and NNX08AO83G. Simulations were carried out at
the National Energy Research Scientific Computing Center.
\end{acknowledgments}


%

\end{document}